\documentclass[aps,prc,reprint,superscriptaddress,showpacs]{revtex4-1}
\usepackage{graphicx}
\usepackage{dcolumn}
\usepackage{multirow}
\usepackage{bm}
\usepackage{epsfig}

\begin{document}

\title{Systematic study of (p,$\gamma$) reactions on Ni isotopes}

\author{A.~Simon}
\email{SimonA@nscl.msu.edu}
\affiliation{National Superconducting Cyclotron Laboratory, Michigan State University, East Lansing, MI 48824, USA}
\affiliation{Joint Institute for Nuclear Astrophysics, Michigan State University, East Lansing, MI 48824, USA}
\author{A.~Spyrou}
\affiliation{National Superconducting Cyclotron Laboratory, Michigan State University, East Lansing, MI 48824, USA}
\affiliation{Joint Institute for Nuclear Astrophysics, Michigan State University, East Lansing, MI 48824, USA}
\affiliation{Department of Physics \& Astronomy, Michigan State University, East Lansing, MI 48824, USA}

\author{T.~Rauscher}
\affiliation{Centre for Astrophysical Research, School of Physics, Astronomy, and Mathematics, University of Hertfordshire, Hatfield AL10 9AB, United Kingdom}
\affiliation{Institute of Nuclear Research (ATOMKI), H-4001 Debrecen, POB 51, Hungary}
\affiliation{Department of Physics, University of Basel, 4056 Basel, Switzerland}

\author{C.~Fr{\"{o}}hlich}
\affiliation{Department of Physics, North Carolina State University, Raleigh, NC 27695, USA}

\author{S.~J.~Quinn}
\affiliation{National Superconducting Cyclotron Laboratory, Michigan State University, East Lansing, MI 48824, USA}
\affiliation{Joint Institute for Nuclear Astrophysics, Michigan State University, East Lansing, MI 48824, USA}
\affiliation{Department of Physics \& Astronomy, Michigan State University, East Lansing, MI 48824, USA}

\author{A.~Battaglia}
\affiliation{Department of Physics and The Joint Institute for Nuclear Astrophysics, University of Notre Dame, Notre Dame, IN 46556, USA}


\author{A.~Best}
\thanks{\emph{Present address:} Lawrence Berkeley National Laboratory, Berkeley, CA 94720, USA}
\affiliation{Department of Physics and The Joint Institute for Nuclear Astrophysics, University of Notre Dame, Notre Dame, IN 46556, USA}

\author{B.~Bucher}
\affiliation{Department of Physics and The Joint Institute for Nuclear Astrophysics, University of Notre Dame, Notre Dame, IN 46556, USA}

\author{M.~Couder}
\affiliation{Department of Physics and The Joint Institute for Nuclear Astrophysics, University of Notre Dame, Notre Dame, IN 46556, USA}

\author{P.~A.~DeYoung}
\affiliation{Department of Physics, Hope College, Holland, MI 49423, USA}

\author{X.~Fang}
\affiliation{Department of Physics and The Joint Institute for Nuclear Astrophysics, University of Notre Dame, Notre Dame, IN 46556, USA}

\author{J.~G\"{o}rres}
\affiliation{Department of Physics and The Joint Institute for Nuclear Astrophysics, University of Notre Dame, Notre Dame, IN 46556, USA}

\author{A.~Kontos}
\thanks{\emph{Present address:} National Superconducting Cyclotron Laboratory, Michigan State University, East Lansing, MI 48824, USA}
\affiliation{Department of Physics and The Joint Institute for Nuclear Astrophysics, University of Notre Dame, Notre Dame, IN 46556, USA}

\author{Q.~Li}
\affiliation{Department of Physics and The Joint Institute for Nuclear Astrophysics, University of Notre Dame, Notre Dame, IN 46556, USA}

\author{L.-Y. Lin}
\affiliation{National Superconducting Cyclotron Laboratory, Michigan State University, East Lansing, MI 48824, USA}
\affiliation{Department of Physics \& Astronomy, Michigan State University, East Lansing, MI 48824, USA}

\author{A.~Long}
\affiliation{Department of Physics and The Joint Institute for Nuclear Astrophysics, University of Notre Dame, Notre Dame, IN 46556, USA}

\author{S.~Lyons}
\affiliation{Department of Physics and The Joint Institute for Nuclear Astrophysics, University of Notre Dame, Notre Dame, IN 46556, USA}



\author{A.~Roberts}
\affiliation{Department of Physics and The Joint Institute for Nuclear Astrophysics, University of Notre Dame, Notre Dame, IN 46556, USA}

\author{D.~Robertson}
\affiliation{Department of Physics and The Joint Institute for Nuclear Astrophysics, University of Notre Dame, Notre Dame, IN 46556, USA}

\author{K.~Smith}
\affiliation{Department of Physics and The Joint Institute for Nuclear Astrophysics, University of Notre Dame, Notre Dame, IN 46556, USA}

\author{M.~K.~Smith}
\affiliation{Department of Physics and The Joint Institute for Nuclear Astrophysics, University of Notre Dame, Notre Dame, IN 46556, USA}

\author{E.~Stech}
\affiliation{Department of Physics and The Joint Institute for Nuclear Astrophysics, University of Notre Dame, Notre Dame, IN 46556, USA}

\author{B.~Stefanek}
\affiliation{National Superconducting Cyclotron Laboratory, Michigan State University, East Lansing, MI 48824, USA}
\affiliation{Department of Physics \& Astronomy, Michigan State University, East Lansing, MI 48824, USA}

\author{W.~P.~Tan}
\affiliation{Department of Physics and The Joint Institute for Nuclear Astrophysics, University of Notre Dame, Notre Dame, IN 46556, USA}

\author{X.~D.~Tang}
\affiliation{Department of Physics and The Joint Institute for Nuclear Astrophysics, University of Notre Dame, Notre Dame, IN 46556, USA}

\author{M.~Wiescher}
\affiliation{Department of Physics and The Joint Institute for Nuclear Astrophysics, University of Notre Dame, Notre Dame, IN 46556, USA}

\date{\today}

\begin{abstract}
A systematic study of the radiative proton capture reaction for all stable nickel isotopes is presented. The results were obtained using 2.0~-~6.0~MeV protons from the 11~MV tandem Van de Graaff accelerator at the University of Notre Dame. The $\gamma$ rays were detected by the NSCL SuN detector utilising the $\gamma$ summing technique. The results are compared to a compilation of earlier measurements and discrepancies between the previous data are resolved.
The experimental results are also compared to the theoretical predictions obtained using the NON-SMOKER and SMARAGD codes. Based on these comparisons an improved set of astrophysical reaction rates is proposed for the (p,$\gamma$) reactions on the stable nickel isotopes as well as for the $^{56}$Ni(p,$\gamma$)$^{57}$Cu reaction.
\end{abstract}


\maketitle

\section{Introduction}
During the past several decades great progress has been made in understanding nucleosynthesis.
Many open questions remain nevertheless, such as the production mechanism of nuclei on the proton rich side of the valley of stability. In particular, there are proton-rich isotopes, the so-called $p$ nuclei, that cannot be produced by neutron capture processes, as they are shielded from $\beta$ decay by the valley of stability \cite{arn03}. A number of processes and sites have been suggested for the production of these nuclides but each has its own problems \cite{raupp11}. The currently favored production process, photodisintegrations in the O/Ne layers of massive stars during their supernova explosion is the so-called $\gamma$ process. This process reproduces the bulk of the $p$ nuclei with the exception of the lightest ones, with mass numbers $A<100$ \cite{woohow,ray95}. Some deficiences have also been found in the region $150\leq A\leq 165$ \cite{rhhw02,hegXX}. Different seed abundances, not encountered in massive stars, may allow a $\gamma$ process to also produce light $p$ nuclides. Thermonuclear explosions of mass accreting White Dwarfs (the single degenerate model for type Ia supernovae) have been suggested as alternative sites but initially no light $p$-nucleus production could be achieved \cite{howmeywoo,howmey,kusa05,kusa11}. Recent simulations, however, found that light $p$ nuclei are produced in sufficient amounts, due to improved hydrodynamic resolution of the nuclear burning zones \cite{travaglio}.

The original work by  \cite{b2fh} suggested to produce proton-rich nuclides in a true $p$ process, $\it{i.\,e.}$, by proton captures in the H-rich envelope of type II supernovae. This was later shown to be unfeasible \cite{autru}. Recently, however, another process was found in the deepest, still ejected layers of a core-collapse supernova, the $\nu p$ process \cite{froh06,pruet,wanajo06}, which involves rapid proton captures on nuclei at and above Ni.
It occurs in explosive environments when proton-rich matter is ejected under the influence of strong neutrino fluxes. When matter in these ejecta expands and cools, nuclear statistical equilibrium mainly comprised of $^4$He, protons and $^{56}$Ni is achieved at temperatures slightly above 4 GK. Rapid proton captures can ensue below about 3.5 GK. Within isotonic chains, (p,$\gamma$)-($\gamma$,p) equilibrium is established and the nuclei with the lowest capture $Q$ values become waiting points where the matter flow through (p,$\gamma$) is halted \cite{raufrohOmeg}. Such nuclei typically also show long $\beta$ decay lifetimes, significantly longer than the expansion timescale.
Without the presence of neutrinos, the matter flow would stop already at $^{64}$Ge \cite{froh06a}.
However, during the explosion the matter is exposed to a large neutrino and antineutrino flux \cite{bur06,thom05}. Due to the high proton abundance, antineutrino captures on free protons can produce free neutrons which allow to bypass the waiting points by (n,p) reactions \cite{froh06a,wanajo11,frohrauOmeg}. The path of the $\nu p$ process is thus defined by the balance between the proton captures and their inverse reactions \cite{frohnupnuc,raufrohOmeg}. Its extension to heavier nuclei depends on the processing speed which is given by the (n,p) reaction rates which, in turn, depend on the neutron abundance generated by the antineutrinos, and the (n,p) reactivities. Accordingly, variation studies have found a strong dependence on uncertainties in (n,p) rates but smaller dependence on (p,$\gamma$) rates because the latter are in equilibrium most of the processing time \cite{wanajo11,frohnupnuc,frohrauOmeg}. Nevertheless, there is a change in the final abundances when varying the proton captures rates because these fall out of equilibrium at the lower end of the $1.5\leq T\leq 3.5$ GK temperature range, within which $\nu p$ processing occurs \cite{wanajo11,froh06a,frohnupnuc,frohrauOmeg}.

All the rates used in astrophysical $\nu p$ process simulations are based on theoretical predictions. It is therefore important to find methods based on experimental data to constrain these rates. The initial part of the $\nu p$-process path follows the $N=Z$ line \cite{wanajo11,raufrohOmeg}. It is closest to stability at the Ni isotopes. 
The goal of this work was to perform a systematic study of the cross sections for (p,$\gamma$) reactions on all stable nickel isotopes, to test the models used for the rate predictions and to provide improved reactivities to be used in the astrophysical simulations. Earlier measurements exist for the discussed isotopes. However, several cases show significant disagreements (up to a factor of five difference in cross section). The current measurements also aimed at resolving those discrepancies.

For each of the targets, the same experimental technique was applied and the same experimental setup was used, to eliminate any uncertainties arising from application of varying detection techniques. All of the results were obtained using the $\gamma$-summing technique.  All the data from previous measurements were retrieved from the EXFOR database \cite{exfor}. The experimental details are presented in Sec.\ \ref{sec:exp}.
New and old data are compared to standard NON-SMOKER \cite{adndt1} calculations as well as to the predictions of the new code SMARAGD v0.9.0s \cite{smaragd} in Sec.\ \ref{sec:theorygeneral}. Based on these results, improved reaction rates for $^{56}$Ni(p,$\gamma$)$^{57}$Cu are suggested in Sec.\ \ref{sec:ni56pg}.

\begin{figure}
\centering
\includegraphics[width=.95\columnwidth]{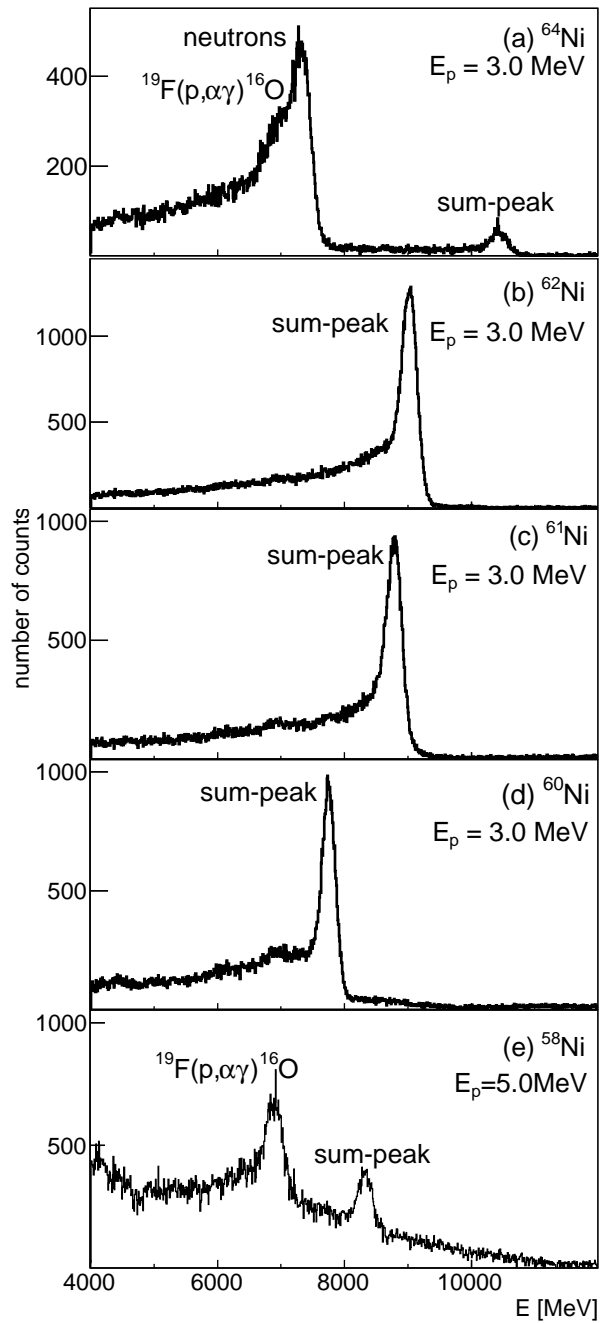}
\caption{\label{fig:ni_spectra}Sample $\gamma$-sum spectra obtained for the (p,$\gamma$) reaction on each of the nickel targets investigated within this work.}
\end{figure}

\section{Experimental details}
\label{sec:exp}
The series of experiments described here was performed using the 11~MV FN tandem Van de Graaff accelerator at the University of Notre Dame.
The targets were irradiated with proton beam in the energy range 2.0~-~6.0~MeV in 0.2~MeV (0.5~MeV in the case of $^{62}$Ni) steps. The measured energy range was chosen to overlap with the previous measurements and when possible extended to cover as much of the Gamow window as possible. The beam intensity was monitored using a Faraday cup.

The thickness of the nickel targets was measured using the Rutherford backscattering technique (RBS) at the Hope College Ion Beam Analysis Laboratory. The properties of the targets are listed in Table \ref{tab:targets}.
Apart from $^{64}$Ni, where the enrichment was 40(5)\%, all the targets were 95(5)\% enriched.

The $\gamma$ rays emitted during the irradiation were detected by the NaI(Tl) segmented summing detector, SuN \cite{sim13}. SuN was developed at the National Superconducting Cyclotron Laboratory, Michigan State University for capture reaction measurements utilizing the $\gamma$-summing technique. The $\gamma$ rays from the decay cascade of the entry state populated during the reaction were summed within the detector. Thus the final spectra were comprised predominantly of the sum peak at the $\gamma$ energy $E_\Sigma=E_\mathrm{c.m.}+Q$, where $E_\mathrm{c.m.}$ is the total kinetic energy in the center of mass system and $Q$ is the reaction $Q$ value. The data was recorded using the NSCL Digital Data Acquisition System (DDAS).

 \begin{table} 
 \caption{\label{tab:targets}Properties of the Ni targets used during the experiment.}
 \begin{ruledtabular}
 \begin{tabular}{lccc}
 Isotope 	& Enrichment 	& Thickness [mg/cm$^2$] 	& $Q$ value [MeV] 	\\
 \hline
 $^{64}$Ni 	& 	40(5)\%		& 0.270(14) 			& 7.452 		\\
 $^{62}$Ni 	& 	95(5)\% 		& 1.66(20) 			& 6.122 		\\
 $^{61}$Ni 	& 	95(5)\% 		& 0.517(67) 			& 5.866 		\\
 $^{60}$Ni 	& 	95(5)\% 		& 0.676(90) 			& 4.800 		\\
 $^{58}$Ni 	& 	95(5)\% 		& 0.943(44) 			& 3.814 		\\

 \end{tabular}
 \end{ruledtabular}
 \end{table}

The data analysis followed the procedure described in detail in \cite{sim13}. For a given beam energy, the sum peak was fitted with a Gaussian combined with a linear background. After subtraction of the background fit, the sum peak was integrated within the region of ($E_\Sigma-3\sigma$,$E_\Sigma+3\sigma$), where $\sigma$ is the standard deviation from the Gaussian fit, to obtain the total number of events. The hit pattern centroid for the events within the same range was used to determine the summing efficiency. 

Sample $\gamma$ spectra obtained for each of the nickel targets are shown in Fig.~\ref{fig:ni_spectra}. As can be seen in Fig. \ref{fig:ni_spectra}, for each reaction the sum peak is clearly observed at the higher energy range of the spectrum. It can be observed in Figs \ref{fig:ni_spectra}(a)-(d), that for a given beam energy, the position of the sum peak changes from target to target reflecting different reaction $Q$ values for various isotopes. The lower energy range ($E<$~4.0~MeV) of the spectrum is dominated by the room background (not shown here). The main background contribution at higher energies comes from the cosmic rays. In two cases ($^{64}$Ni and $^{58}$Ni), fluorine contamination in the target resulted in a peak at 6.92 and 7.12~MeV from the $^{19}$F(p,$\alpha\gamma$)$^{16}$O reaction. This contamination limited the energy range accessible for the measurements with the $^{58}$Ni target, as for the beam energy below 4.4~MeV the sum-peak energy ($E_\Sigma=E_\mathrm{c.m.}+Q$) would overlap with the contamination lines. Additionally, in Fig. \ref{fig:ni_spectra}(a) a peak at 7.2~MeV is present that can be related to the neutrons from the (p,n) reaction in the target.

During the analysis, contributions from other isotopes were also considered. In particular, for the lightest nickel isotopes the (p,$\gamma$) reaction cross section is an order of magnitude lower than for heavier isotopes, thus even a small concentration of the heavier isotopes could result in a non-negligible sum peak. However, no such contributions were found. 

The (p,$\gamma$) reaction cross sections of all stable nickel isotopes obtained in this work are listed in Table \ref{tab:cs}.
The uncertainty of the cross section values includes the statistical uncertainty as well as the uncertainty resulting from subtraction of the background under the sum peak. The uncertainty in the target thickness of 5\% was also included, as well as the uncertainty of the summing efficiency obtained using the method described in \cite{sim13} (typically of the order of 10\% relative uncertainty).

 \begin{table} 
 \caption{\label{tab:cs}Cross section for the (p,$\gamma$) reactions obtained in this work for all stable nickel isotopes.}
 \begin{ruledtabular}
 \begin{tabular}{cccc}
 E$_\mathrm{c.m.}$ [MeV] 	& $\sigma$ [mb] &  E$_\mathrm{c.m.}$ [MeV] 	& $\sigma$ [mb]\\
 \hline
 \multicolumn{4}{c}{$^{64}$Ni(p,$\gamma$)$^{65}$Cu}\\
 \hline
1.96&	0.66(11)	&	3.54	&	0.12(02)	\\
2.16&	1.13(18)	&	3.73	&	0.20(03)	\\
2.35&	1.39(22)	&	3.93	&	0.17(03)	\\
2.55&	1.03(16)	&	4.14	&	0.26(04)	\\
2.75	&	0.33(05)	&	4.33	&	0.22(04)	\\
2.94	&	0.15(02)	&	4.52	&	0.30(05)	\\
3.14	&	0.24(04)	&	4.72	&	0.22(04)	\\
3.34	&	0.15(02)	&	4.92	&	0.28(05)	\\
 \hline
 \multicolumn{4}{c}{$^{62}$Ni(p,$\gamma$)$^{63}$Cu}\\
 \hline
2.41&	1.08(14)		&	3.40	&	1.99(25)	\\
2.90	&	1.379(17)	&	3.93	&	2.73(34)	\\
 \hline
 \multicolumn{4}{c}{$^{61}$Ni(p,$\gamma$)$^{62}$Cu}\\
 \hline
1.95	&	0.34(04)	&	2.74	&	0.92(11)	\\
2.14	&	0.49(06)	&	2.84	&	0.89(11)	\\
2.34	&	0.64(08)	&	2.94	&	0.96(12)	\\
2.54	&	0.75(09)	&	3.03	&	0.83(10)	\\
 \hline
 \multicolumn{4}{c}{$^{60}$Ni(p,$\gamma$)$^{61}$Cu}\\
 \hline
2.73		&	0.30(04)		&	4.31		&	0.97(12)\\
2.93		&	0.27(03)		&	4.51		&	1.37(16)	\\
3.13		&	0.25(03)		&	4.71		&	0.87(11)	\\
3.33		&	0.50(06)		&	4.90		&	0.91(12)	\\
3.52		&	0.42(05)		&	5.10		&	2.37(30)	\\
3.72		&	0.65(08)		&	5.49		&	2.68(34)	\\
3.92		&	0.70(09)		&	5.89		&	2.32(30)	\\
4.11		&	1.01(13)		&	&\\
 \hline
 \multicolumn{4}{c}{$^{58}$Ni(p,$\gamma$)$^{59}$Cu}\\
 \hline
4.30		&0.147(19)	&5.09		&0.266(34)\\
4.50		&0.161(20)	&5.29		&0.432(56)\\
4.70		&0.134(17)	&5.49		&0.482(62)\\
4.89		&0.152(19)	&5.88		&0.378(49)\\
 \end{tabular}
 \end{ruledtabular}
 \end{table}

\section{Theoretical analysis -- general considerations}
\label{sec:theorygeneral}

\begin{figure}
\includegraphics[angle=-90,width=\columnwidth]{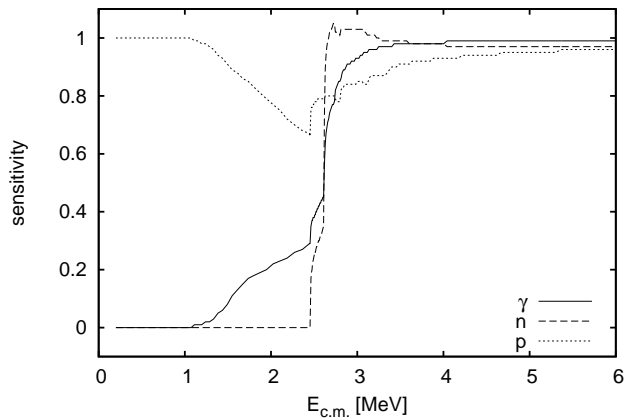}
\caption{\label{fig:sensiNi64}Sensitivity of the reaction cross sections of $^{64}$Ni(p,$\gamma$)$^{65}$Cu to variations of $\gamma$ and particle widths.}
\end{figure}

\begin{figure}
\includegraphics[angle=-90,width=\columnwidth]{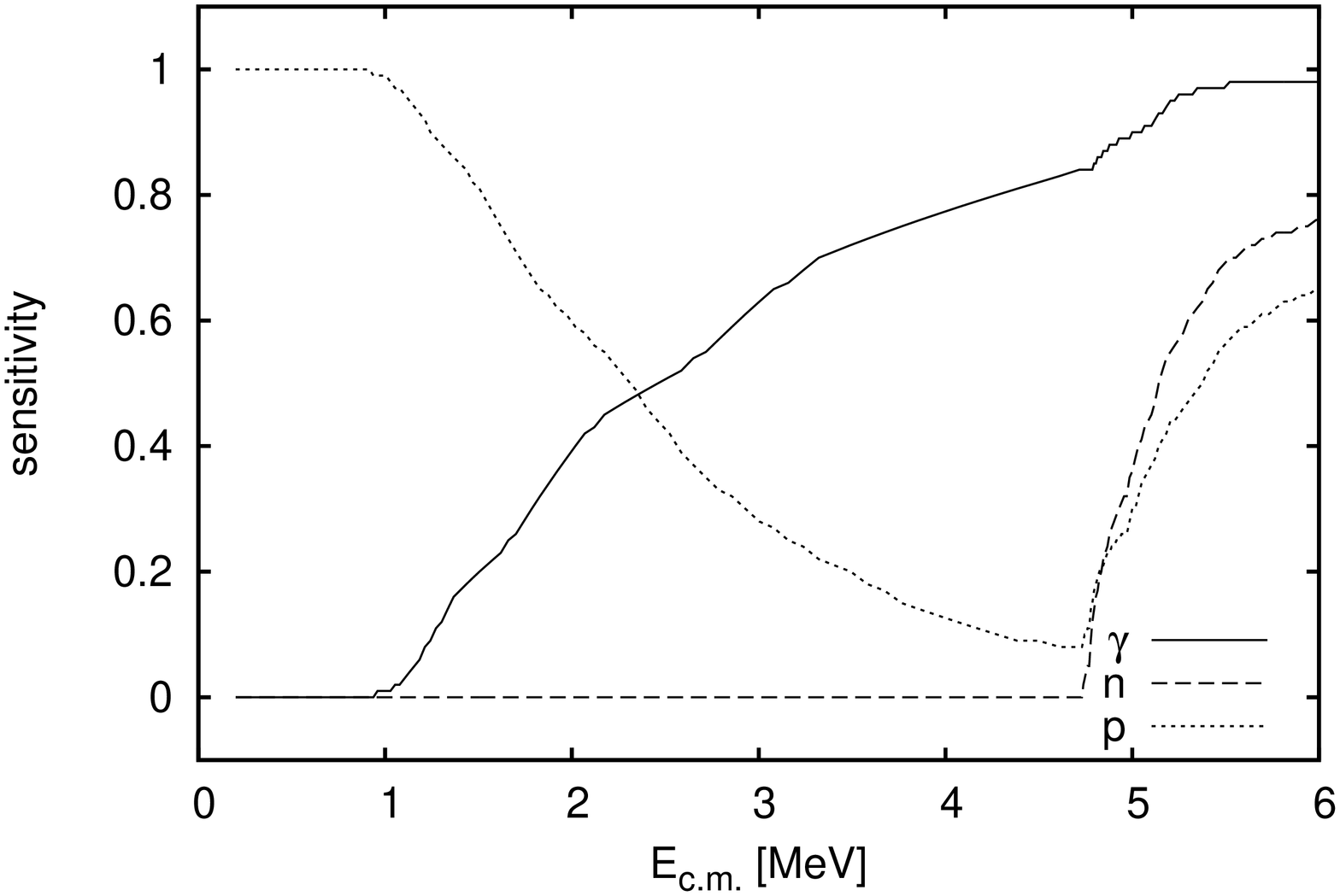}
\caption{\label{fig:sensiNi62}Sensitivity of the reaction cross sections of $^{62}$Ni(p,$\gamma$)$^{63}$Cu to variations of $\gamma$ and particle widths.}
\end{figure}

\begin{figure}
\includegraphics[angle=-90,width=\columnwidth]{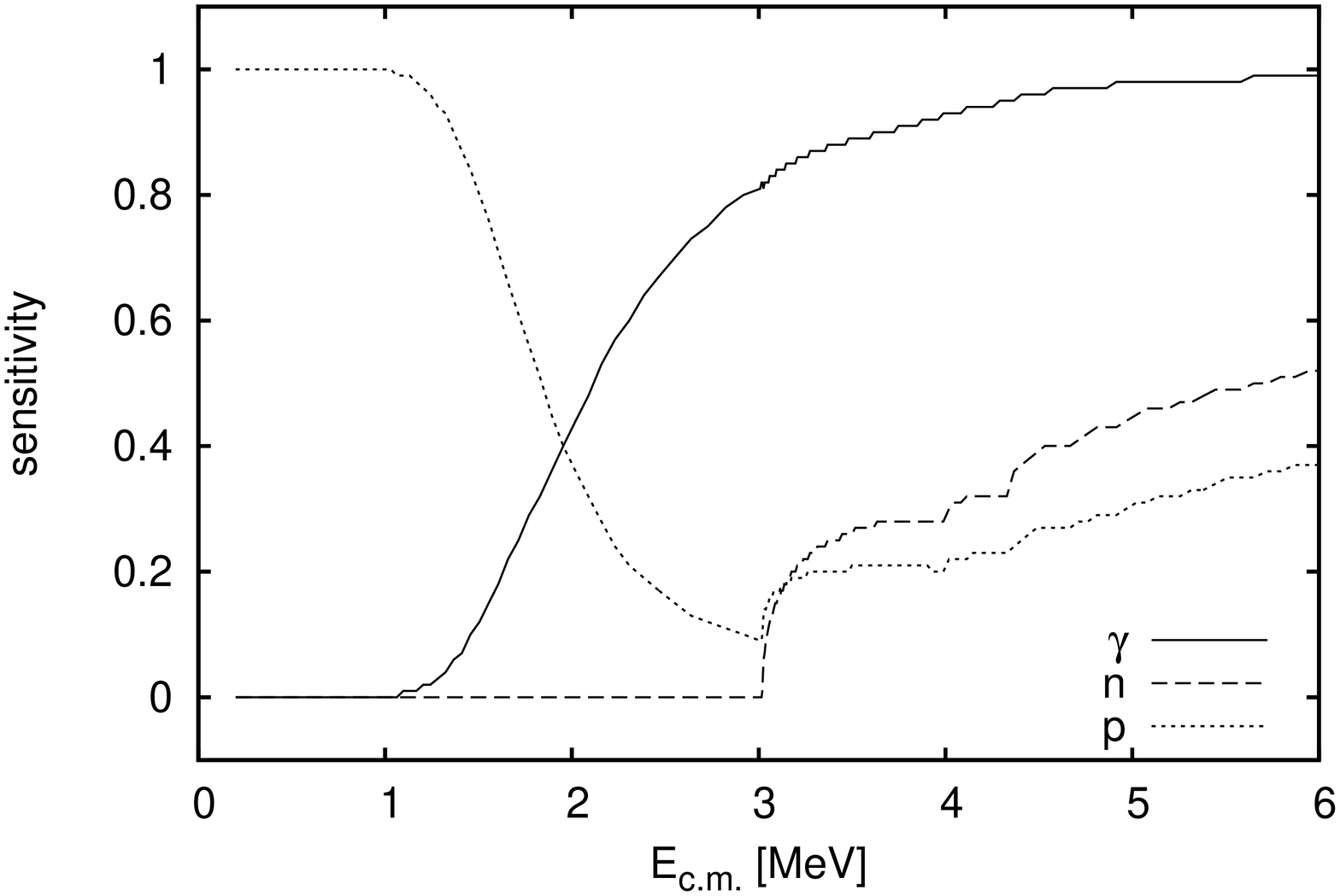}
\caption{\label{fig:sensiNi61}Sensitivity of the reaction cross sections of $^{61}$Ni(p,$\gamma$)$^{62}$Cu to variations of $\gamma$ and particle widths.}
\end{figure}

\begin{figure}
\includegraphics[angle=-90,width=\columnwidth]{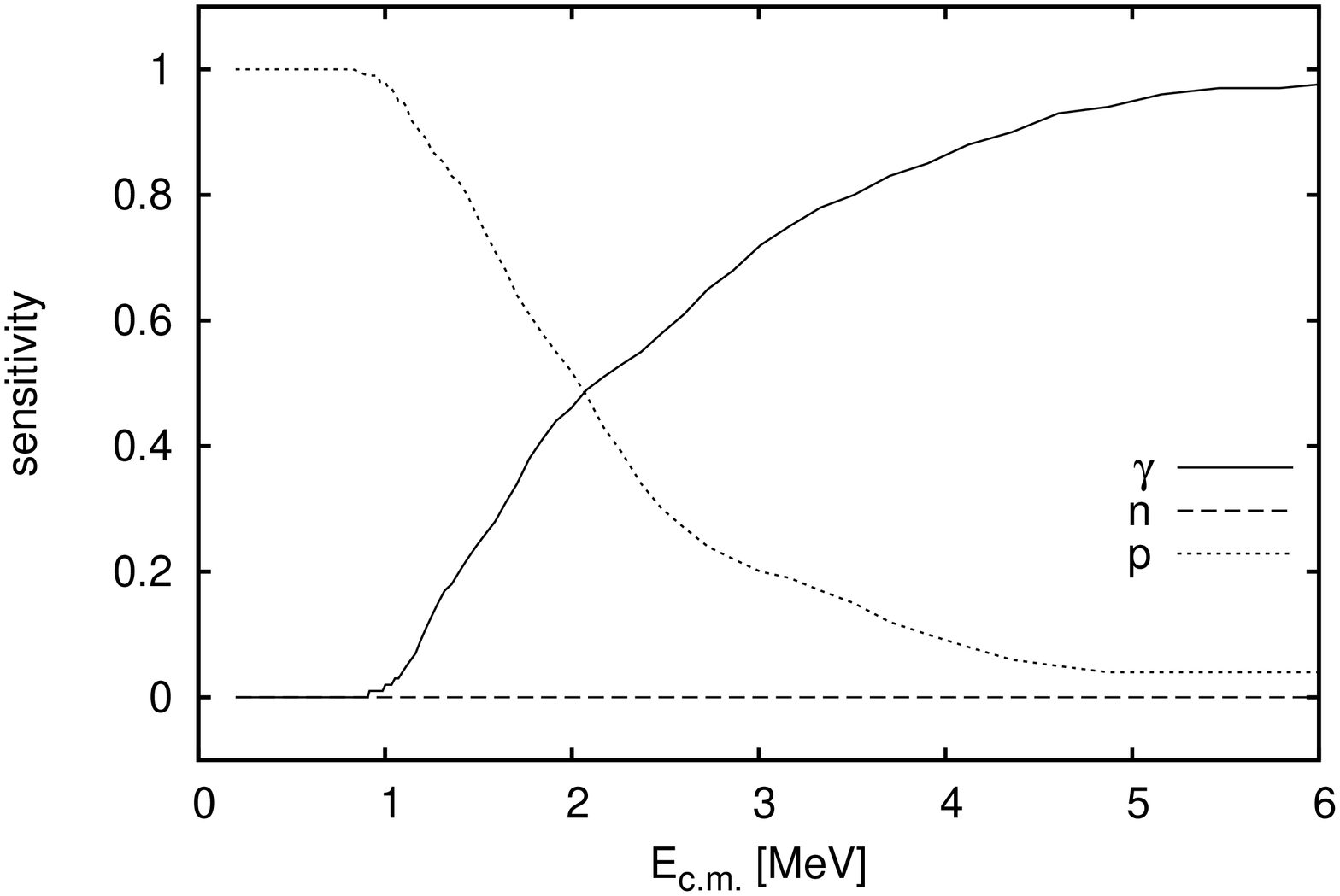}
\caption{\label{fig:sensiNi60}Sensitivity of the reaction cross sections of $^{60}$Ni(p,$\gamma$)$^{61}$Cu to variations of $\gamma$ and particle widths.}
\end{figure}

\begin{figure}
\includegraphics[angle=-90,width=.95\columnwidth]{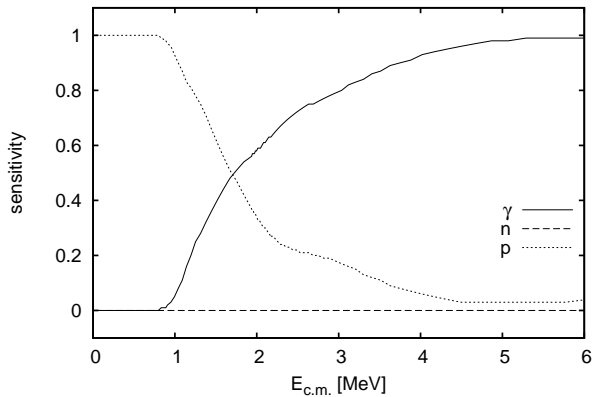}
\caption{\label{fig:sensiNi58}Sensitivity of the reaction cross sections of $^{58}$Ni(p,$\gamma$)$^{59}$Cu to variations of $\gamma$ and particle widths.}
\end{figure}

The studied reactions proceed via resonant compound formation, either through isolated resonances or a high number of overlapping ones, and thus depend on particle and radiation widths. From the comparisons of the experimental results with the theoretical predictions it will become evident that each case has to be analyzed separately because the dependence of the cross sections on the various widths is different for each case. A simple comparison of data and predictions without considering these dependencies will be misleading and may not be able to identify the underlying deficiency trends and thus also the impact on the astrophysical reaction rates correctly. At temperatures relevant for the $\nu p$ process the reaction rates mainly depend on the proton widths. The reaction rate sensitivity on the $\gamma$ width increases with decreasing neutron number of the compound nucleus as its proton separation energy $S_\mathrm{p}$ is decreasing. Therefore the knowledge of the optical proton+nucleus potential, required for the calculation of the proton width, is essential to constrain the rate in all cases.
All sensitivities and their definition can be found in \cite{sensi}. Figures \ref{fig:sensiNi64}-\ref{fig:sensiNi58} show the cross section sensitivities required in the analyses of the present work. All cross sections are insensitive to variations of $\alpha$-widths in the shown energy range.

The data are compared to calculations within the statistical Hauser-Feshbach model \cite{hf}. This model assumes that a large number of resonances is present at the compound formation energy and that their individual widths can be replaced by average widths. Required is a ``sufficiently'' high nuclear level density (NLD) at the compound formation energy $E_\mathrm{c}=S_\mathrm{p}+E_\mathrm{c.m.}$ \cite{hwfz,rtk,raureview}. Except for $^{61}$Cu and $^{59}$Cu, all NLDs are large enough at $E_\mathrm{c}$ to apply the statistical model. The NLD is clearly too low in $^{58}$Ni(p,$\gamma$)$^{59}$Cu, and $^{60}$Ni(p,$\gamma$)$^{61}$Cu seems to be at the borderline. Even when the model is not applicable, i.e., when resonance structures show up in the excitation functions, the statistical model should still be able to give an average value for the cross sections. If several resonances are contributing, this average value may be sufficient for the computation of the reaction rate which involves an integration over the cross sections.

Finally, it has to be considered whether stellar rates can directly be inferred from the data or whether additional theory has to be invoked. This depends on the contribution of reactions commencing on thermally populated, excited states of the target nuclei in the stellar plasma. A ground state (g.s.) contribution, i.e., the contribution of the laboratory cross section to the stellar one, of $X_0 \simeq 1$ is required to ensure that excited states do not contribute \cite{fow74,xfactor,rauapjlett}. This is fulfilled for all investigated reactions, except for $^{61}$Ni(p,$\gamma$)$^{62}$Cu which shows $X_0=0.52$ at 1.5 GK \cite{sensi}.

In the following, the reactions are discussed one by one, starting with $^{64}$Ni(p,$\gamma$)$^{65}$Cu and going towards the more neutron-deficient target nuclei. The cross section predictions and width variation studies were performed with the code SMARAGD, version 0.9.0s \cite{raureview,smaragd}. Additionally, comparisons to the standard NON-SMOKER cross sections and reaction rates \cite{adndt1,adndt} are shown. The standard optical proton potential used in those calculations is from \cite{lej}, which is a low-energy modification of the potential by \cite{jlm} and was provided especially for astrophysical applications. The potential of \cite{jlm} uses the Br\"uckner-Hartree-Fock approximation with Reid's hard core
nucleon-nucleon interaction and adopts a local density approximation. Reparameterizations of \cite{jlm} using more recent data are \cite{bauge} and its Lane-consistent version \cite{baugelane}. They do not, however, adopt special considerations for low proton energies.

Another modification of \cite{lej}, using an increased imaginary part, was developed for an improved description of low-energy (p,$\gamma$) and (p,n) reactions on intermediate nuclei \cite{raureview,cdpg,kisssupp,tomsupp}. The present data and calculations are not sufficient to distinguish between the two versions with and without the modified imaginary part. Due to the uncertainties in the description of the widths, similar results can be obtained using either potential. Therefore only calculations using \cite{lej} are shown here.

\section{Results}

\subsection{$^{64}$Ni(p,$\gamma$)$^{65}$Cu}

\begin{figure}
\includegraphics[angle=-90,width=\columnwidth]{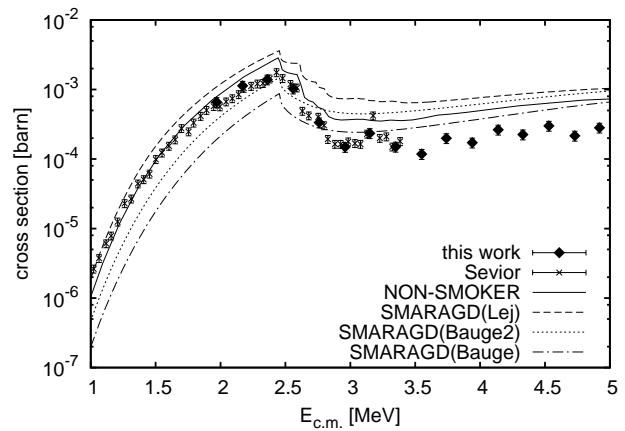}
\caption{\label{fig:ni64comp}Comparison of measured reaction cross sections of $^{64}$Ni(p,$\gamma$)$^{65}$Cu to predictions: NON-SMOKER \cite{adndt,adndt1} (using the potential of \cite{lej}) and SMARAGD with the potentials of \cite{lej} (Lej), \cite{baugelane} (Bauge), and \cite{bauge} (Bauge2). The data are from this work and from \cite{sevior} (Sevior).}
\end{figure}

The new results agree well with the previous measurement by \cite{sevior}.
As shown in Fig.\ \ref{fig:sensiNi64}, the reaction cross sections are sensitive to all widths above the (p,n) threshold (2.495~MeV) while they quickly become only dependent on the proton width when going to lower energies. A comparison with predictions using various optical proton+nucleus potentials (and with the standard NON-SMOKER prediction) is shown in Fig.\ \ref{fig:ni64comp}. The SMARAGD calculation with the potential by \cite{lej} reproduces the data at the lower edge of the energy range in which measurements are available but diverges when going to higher energy. Our new data allow a detailed study of the cause of this divergence because they extend to higher energy.

\begin{figure}
\includegraphics[angle=-90,width=\columnwidth]{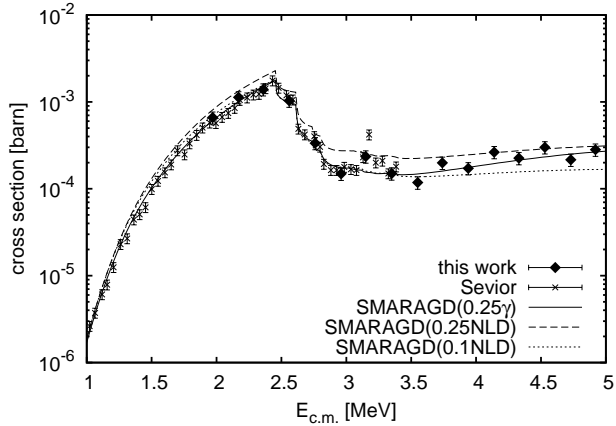}
\caption{\label{fig:ni64solution}Comparison of measured reaction cross sections of $^{64}$Ni(p,$\gamma$)$^{65}$Cu to predictions using a $\gamma$ width divided by a factor of 4 ($0.25\gamma$), and NLDs divided by 4 (0.25NLD) and 10 (0.1NLD), respectively. The data are from this work and from \cite{sevior} (Sevior).}
\end{figure}

A rescaling of the proton width cannot lead to an improvement in the predictions as it would affect the cross sections at all energies.
Since the neutron width only impacts the cross sections above the (p,n) threshold and the deviations also occur below it, the only possibility is a change in the $\gamma$ width. Figure \ref{fig:ni64solution} shows how a calculation with the standard potential by \cite{lej} but a $\gamma$ width divided by 4 reproduces the data well across the full measured energy range.

The $\gamma$ width, in turn, has three dependencies: on the low-lying discrete excited states of the compound nucleus included in the calculation, on the NLD above the last included excited state, and on the $\gamma$-strength function. Thanks to the new data it is possible to distinguish between the three possibilities. The major contribution to the $\gamma$ width comes from $\gamma$ transitions leading to states about 2~-~4~MeV below the compound formation energy \cite{raugamma}. Since the proton separation energy in $^{65}$Cu is larger than these $\gamma$ energies, the low-lying, discrete states only play a minor role in this case. A variation of the NLD yields different changes in the dependence than varying the $\gamma$ width in total, as shown in Fig.\ \ref{fig:ni64solution}. Therefore the deficiency in the predicted $\gamma$ width stems from the strength function used, i.e., the GDR description. This is inconsequential for the astrophysical rate, as it is only sensitive to the proton width.

Within the measured energy range, the experimental data is well reproduced by the SMARAGD calculations with the renormalized $\gamma$ width but unchanged proton width. Adopting this prediction also in the astrophysically relevant energy range of 0.96~-~2.52~MeV \cite{energywindows} leads to a higher rate than the one of the standard NON-SMOKER prediction. The new reactivity is given in Table \ref{tab:nirate}. The parameters of a fit in the REACLIB format are given in Table \ref{tab:nifit}. The reactivities across a wider temperature range are provided as this is necessary for using the table directly in nucleosynthesis reaction networks. It also enables users to make their own fit, if they so desire. The data only constrain the rates above 5 GK but this is of no consequence for reaction networks, as above 5 GK all rates are in equilibrium and rate values only have to be provided to avoid divisions by zero.

\begin{table*}
\caption{Predicted reactivities of (p,$\gamma$) reactions on all stable nickel isotopes.\label{tab:nirate}}
\begin{ruledtabular} \begin{tabular}{clllll}
\multirow{2}{*}{$T$ [GK]}		& \multicolumn{5}{c}{$N_\mathrm{A} \left< \sigma v \right>$ [cm$^3$ s$^{-1}$ mole$^{-1}$]}\\
\cline{2-6}
 		& $^{64}$Ni 	& $^{62}$Ni 		& $^{61}$Ni 		& $^{60}$Ni 		& $^{58}$Ni \\
\hline
0.1	&	$5.846\times 10^{-20}$   &	$3.991\times 10^{-20}$  &	$3.534\times 10^{-20}$  &	$2.058\times 10^{-20}$  &	$1.240\times 10^{-20}$  \\
0.15	&	$2.007\times 10^{-15}$   &	$1.372\times 10^{-15}$  &	$1.215\times 10^{-15}$  &	$7.041\times 10^{-16}$  &	$4.213\times 10^{-16}$  \\
0.2	&	$1.359\times 10^{-12}$   &	$9.316\times 10^{-13}$  &	$8.252\times 10^{-13}$  &	$4.763\times 10^{-13}$  &	$2.832\times 10^{-13}$  \\
0.3	&	$4.728\times 10^{-9}$   &	$3.260\times 10^{-9}$  &	$2.891\times 10^{-9}$  &	$1.657\times 10^{-9}$  &	$9.744\times 10^{-10}$  \\
0.4	&	$7.984\times 10^{-7}$   &	$5.542\times 10^{-7}$  &	$4.922\times 10^{-7}$  &	$2.791\times 10^{-7}$  &	$1.621\times 10^{-7}$  \\
0.5	&	$3.031\times 10^{-5}$   &	$2.118\times 10^{-5}$  &	$1.883\times 10^{-5}$  &	$1.047\times 10^{-5}$  &	$5.930\times 10^{-6}$  \\
0.6	&	$4.793\times 10^{-4}$   &	$3.361\times 10^{-4}$  &	$2.989\times 10^{-4}$  &	$1.614\times 10^{-4}$  &	$8.749\times 10^{-5}$  \\
0.7	&	$4.274\times 10^{-3}$   &	$2.999\times 10^{-3}$  &	$2.665\times 10^{-3}$  &	$1.385\times 10^{-3}$  &	$7.092\times 10^{-4}$  \\
0.8	&	$2.548\times 10^{-2}$   &	$1.783\times 10^{-2}$  &	$1.582\times 10^{-2}$  &	$7.860\times 10^{-3}$  &	$3.789\times 10^{-3}$  \\
0.9	&	$1.128\times 10^{-1}$   &	$7.861\times 10^{-2}$  &	$6.959\times 10^{-2}$  &	$3.294\times 10^{-2}$  &	$1.497\times 10^{-2}$  \\
1.0	&	$3.986\times 10^{-1}$   &	$2.761\times 10^{-1}$  &	$2.436\times 10^{-1}$  &	$1.099\times 10^{-1}$  &	$4.725\times 10^{-2}$  \\
1.5	&	$2.809\times 10^{1}$     &	$1.887\times 10^{1}$    &	$1.617\times 10^{1}$    &	5.973    &	2.072    \\
2.0	&	$3.314\times 10^{2}$     &	$2.199\times 10^{2}$    &	$1.803\times 10^{2}$    &	$5.900\times 10^{1}$    &	$1.780\times 10^{1}$    \\
2.5	&	$1.653\times 10^{3}$     &	$1.127\times 10^{3}$    &	$8.745\times 10^{2}$    &	$2.689\times 10^{2}$    &	$7.363\times 10^{1}$    \\
3.0	&	$5.005\times 10^{3}$     &	$3.669\times 10^{3}$    &	$2.660\times 10^{3}$    &	$8.052\times 10^{2}$    &	$2.051\times 10^{2}$    \\
3.5	&	$1.100\times 10^{4}$     &	$9.013\times 10^{3}$    &	$6.025\times 10^{3}$    &	$1.868\times 10^{3}$    &	$4.487\times 10^{2}$    \\
4.0	&	$1.941\times 10^{4}$     &	$1.830\times 10^{4}$    &	$1.115\times 10^{4}$    &	$3.666\times 10^{3}$    &	$8.355\times 10^{2}$    \\
4.5	&	$2.925\times 10^{4}$     &	$3.233\times 10^{4}$    &	$1.782\times 10^{4}$    &	$6.388\times 10^{3}$    &	$1.385\times 10^{3}$    \\
5.0	&	$3.915\times 10^{4}$     &	$5.120\times 10^{4}$    &	$2.550\times 10^{4}$    &	$1.017\times 10^{4}$    &	$2.100\times 10^{3}$    \\
6.0	&	$5.410\times 10^{4}$     &	$9.952\times 10^{4}$    &	$4.099\times 10^{4}$    &	$2.077\times 10^{4}$    &	$3.924\times 10^{3}$    \\
7.0	&	$5.814\times 10^{4}$     &	$1.486\times 10^{5}$    &	$5.241\times 10^{4}$    &	$3.337\times 10^{4}$    &	$5.913\times 10^{3}$    \\
8.0	&	$5.195\times 10^{4}$     &	$1.811\times 10^{5}$    &	$5.717\times 10^{4}$    &	$4.359\times 10^{4}$    &	$7.547\times 10^{3}$    \\
9.0	&	$4.052\times 10^{4}$     &	$1.876\times 10^{5}$    &	$5.565\times 10^{4}$    &	$4.798\times 10^{4}$    &	$8.468\times 10^{3}$    \\
10.0	&	$2.895\times 10^{4}$	 &	$1.707\times 10^{5}$   &	$4.987\times 10^{4}$    &	$4.623\times 10^{4}$    &	$8.596\times 10^{3}$

\end{tabular} \end{ruledtabular}
\end{table*}

\begin{table*}
\caption{Fit parameters (in REACLIB format \cite{adndt}) for the reactivity of (p,$\gamma$) and its reverse reaction for all stable nickel isotopes. The reverse value has to be multiplied by the ratio of the partition functions to obtain the ($\gamma$,p) reactivity.\label{tab:nifit}}
\begin{ruledtabular} \begin{tabular}{clllllll}
&$a_0$ 		& $a_1$ 			&$a_2$ 			& $a_3$ 			& $a_4$ 		& $a_5$ 			& $a_6$\\
\hline
&\multicolumn{7}{c}{$^{64}$Ni}\\
\hline
&  $7.414078\times 10^{1}$& $-1.175453$&  $2.162346\times 10^{1}$& $-1.003880\times 10^{2}$&  $5.252775$& $-3.743412\times 10^{-1}$&  $4.653832\times 10^{1}$\\
rev&  $9.643699\times 10^{1}$& $-8.766878\times 10^{1}$&  $2.162346\times 10^{1}$& $-1.003880\times 10^{2}$&  $5.252775$& $-3.743412\times 10^{-1}$&  $4.803832\times 10^{1}$\\
\hline
&\multicolumn{7}{c}{$^{62}$Ni}\\
\hline
&  $1.173687\times 10^{2}$& $-2.367050$&  $8.350530\times 10^{1}$& $-2.120871\times 10^{2}$&  $1.315014\times 10^{1}$& $-8.584441\times 10^{-1}$&  $9.604393\times 10^{1}$\\
rev &  $1.396641\times 10^{2}$& $-7.341479\times 10^{1}$&  $8.350530\times 10^{1}$& $-2.120871\times 10^{2}$&  $1.315014\times 10^{1}$& $-8.584441\times 10^{-1}$&  $9.754393\times 10^{1}$\\
\hline
&\multicolumn{7}{c}{$^{61}$Ni}\\
\hline
&  $9.846000\times 10^{1}$& $-1.722409$&  $5.297264\times 10^{1}$& $-1.598366\times 10^{2}$&  $9.315694$& $-6.041871\times 10^{-1}$&  $7.246493\times 10^{1}$\\
rev &  $1.217359\times 10^{2}$& $-6.979359\times 10^{1}$&  $5.297264\times 10^{1}$& $-1.598366\times 10^{2}$&  $9.315694$& $-6.041871\times 10^{-1}$&  $7.396493\times 10^{1}$\\
\hline
&\multicolumn{7}{c}{$^{60}$Ni}\\
\hline
&  $1.306414\times 10^{2}$& $-1.868247$&  $7.166889\times 10^{1}$& $-2.159482\times 10^{2}$&  $1.421891\times 10^{1}$& $-9.256727\times 10^{-1}$&  $9.242814\times 10^{1}$ \\
rev&  $1.529360\times 10^{2}$& $-5.757600\times 10^{1}$&  $7.166889\times 10^{1}$& $-2.159482\times 10^{2}$&  $1.421891\times 10^{1}$& $-9.256727\times 10^{-1}$&  $9.392814\times 10^{1}$\\
\hline
&\multicolumn{7}{c}{$^{58}$Ni}\\
\hline
&  $1.025240\times 10^{2}$&  0.000000& $-1.002597\times 10^{1}$& $-1.027998\times 10^{2}$&  $7.767831$& $-5.337708\times 10^{-1}$&  $3.462199\times 10^{1}$ \\
rev &  $1.248178\times 10^{2}$& $-3.967033\times 10^{1}$& $-1.002597\times 10^{1}$& $-1.027998\times 10^{2}$&  $7.767831$& $-5.337708\times 10^{-1}$&  $3.612199\times 10^{1}$
\end{tabular} \end{ruledtabular}
\end{table*}

\subsection{$^{62}$Ni(p,$\gamma$)$^{63}$Cu}

\begin{figure}
\includegraphics[angle=-90,width=\columnwidth]{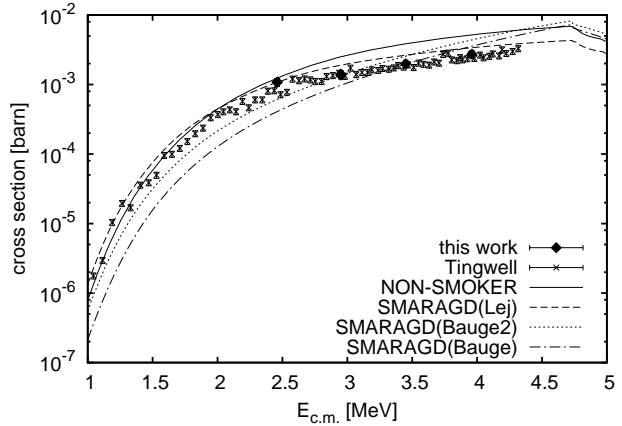}
\caption{\label{fig:ni62comp}Same as Fig.\ \ref{fig:ni64comp} but for $^{62}$Ni(p,$\gamma$)$^{63}$Cu. The data are from this work and from \cite{tin88} (Tingwell).}
\end{figure}

The cross section sensitivities of $^{62}$Ni(p,$\gamma$)$^{63}$Cu are quite similar to those of $^{64}$Ni(p,$\gamma$)$^{65}$Cu but the sensitivity to the $\gamma$ width is more slowly declining towards lower energies. Contrary to the situation with $^{64}$Ni(p,$\gamma$)$^{65}$Cu, the SMARAGD calculation using the standard proton potential by \cite{lej} is in good agreement with the data across the measured energy range. The new data obtained within this work confirm the previous measurements \cite{tin88}, as can be seen in Fig.\ \ref{fig:ni62comp}. At the high energy end, there is a slightly larger deviation between theory and data but it is only 10~-~20\% which is within the expected range of accuracy of a global prediction. Again, there is good agreement at the lower end of the measured range, where the proton width is determining the cross section.

The stellar reactivity of the standard SMARAGD prediction is given in Table \ref{tab:nirate} and the fit parameters are listed in Table \ref{tab:nifit}.

\subsection{$^{61}$Ni(p,$\gamma$)$^{62}$Cu}

\begin{figure}
\includegraphics[angle=-90,width=\columnwidth]{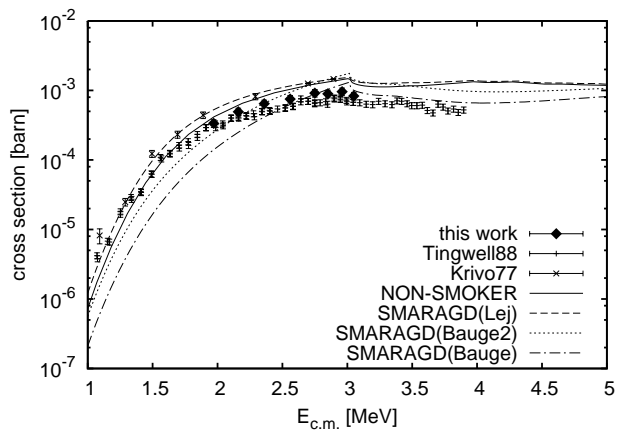}
\caption{\label{fig:ni61comp}Same as Fig.\ \ref{fig:ni64comp} but for $^{61}$Ni(p,$\gamma$)$^{62}$Cu. The data are from this work and from \cite{kriv77} (Krivo77), \cite{tin88a} (Tingwell88).}
\end{figure}


With decreasing neutron number, the cross sections above the (p,n) threshold depend less on the neutron- and proton-widths and primarily depend on the $\gamma$ width. Below the threshold, its importance decreases again and the proton width determines the cross sections in the astrophysical energy range. This is also the case for $^{61}$Ni.

\begin{figure}
\includegraphics[angle=-90,width=\columnwidth]{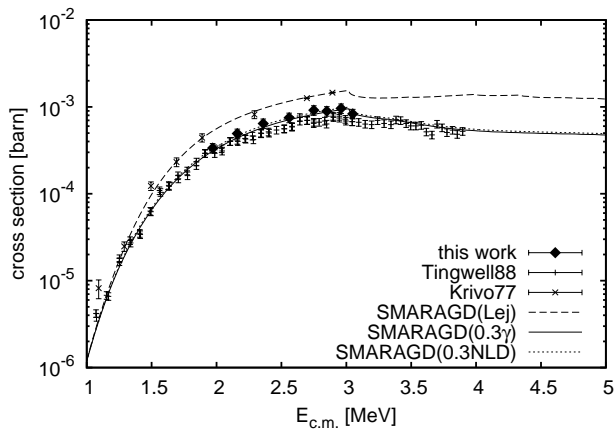}
\caption{\label{fig:ni61nld}Comparison of measured reaction cross sections of $^{61}$Ni(p,$\gamma$)$^{62}$Cu to predictions using a $\gamma$ width multiplied by a factor of 0.3 ($0.3\gamma$) and NLDs multiplied by the same factor (0.3NLD). The data are from this work and from \cite{kriv77} (Krivo77), \cite{tin88a} (Tingwell88).}
\end{figure}

Figure \ref{fig:ni61comp} shows the comparison of various predictions to the data, similar as in the previous sections. Two data sets from \cite{kriv77,tin88a} for $^{61}$Ni(p,$\gamma$)$^{62}$Cu reaction cross section were found in the literature, and a significant discrepancy between these data exists. Good agreement between the current work and the results from \cite{tin88a} was found. The SMARAGD calculation using the standard potential by \cite{lej} describes the data of \cite{kriv77} well. The data of \cite{tin88a} and those obtained within this work can be reproduced by reducing the $\gamma$ width by a factor of 0.3, see Fig.\ \ref{fig:ni61nld}. Contrary to the $^{64}$Ni(p,$\gamma$)$^{65}$Cu case, however, it is impossible to decide the reason for the reduction. As shown in Fig.\ \ref{fig:ni61nld}, the variation in the NLD leads to indistinguishable results from the variation of the full $\gamma$ width. A simple rescaling of the $\gamma$ strength would lead to the same result.

The comparison of the SMARAGD prediction is compatible with an unchanged proton width and therefore we can assume that the astrophysical reactivity is also predicted well. It is given in Table \ref{tab:nirate} and the REACLIB fit coefficients are in Table \ref{tab:nifit}.

\subsection{$^{60}$Ni(p,$\gamma$)$^{61}$Cu}

\begin{figure}
\includegraphics[angle=-90,width=\columnwidth]{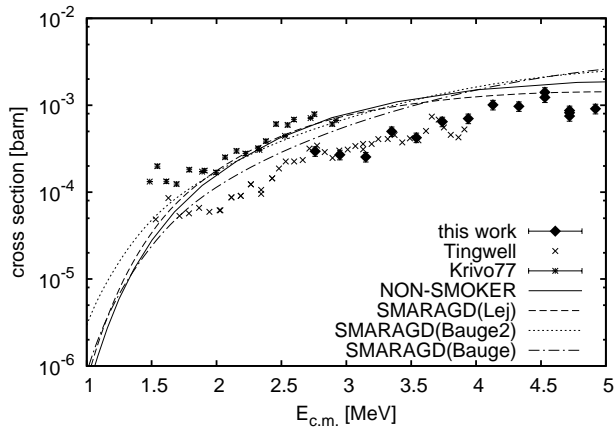}
\caption{\label{fig:ni60comp}Same as Fig.\ \ref{fig:ni64comp} but for $^{60}$Ni(p,$\gamma$)$^{61}$Cu. Data are from this work and from \cite{kriv77} (Krivo77), \cite{tin88} (Tingwell).}
\end{figure}

The proton separation energy in $^{61}$Cu is only 4.801~MeV whereas the neutron separation energy is 11.711~MeV. The data compared to predictions in Fig.\ \ref{fig:ni60comp} do not reach the (p,n) threshold and also only barely reach the region where the proton width is dominating at low energy. The reaction rate would be equally sensitive to proton and $\gamma$ width at the high temperature of 3.5 GK but rapidly becomes sensitive to only the proton width towards lower temperature, including the $\nu p$ process freeze-out temperature of 1.5 GK.

In the case of the $^{60}$Ni(p,$\gamma$)$^{61}$Cu reaction, the two data sets \cite{kriv77,tin88} found in literature do not agree with each other. The results of this work, as can be seen in Fig. \ref{fig:ni60comp}, are in a good agreement with those of \cite{tin88} and extend to higher energies, allowing to better constrain the energy dependence. The SMARAGD calculation is close to the new data at the upper end of the measured range but is closer to the result of \cite{kriv77} otherwise.

\begin{figure}
\includegraphics[angle=-90,width=\columnwidth]{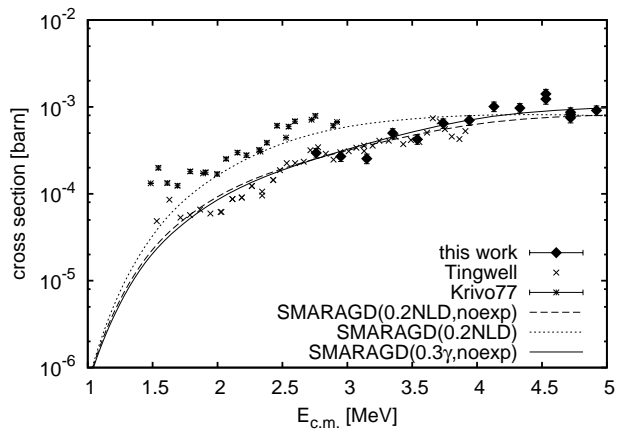}
\caption{\label{fig:ni60nld}Comparison of measured reaction cross sections of $^{60}$Ni(p,$\gamma$)$^{61}$Cu to predictions using NLDs divided by 5, with (0.2NLD) and without (0.2NLD,noexp) excited states for the compound nucleus included in the calculation. Data are from this work and from \cite{kriv77} (Krivo77), \cite{tin88} (Tingwell).}
\end{figure}

It proved difficult to obtain the energy dependence of the combined data of our measurement and the one by \cite{tin88}. It is only possible by using a scaled NLD \textit{without} including experimentally known excited states in the compound nucleus. Figure \ref{fig:ni60nld} shows the theoretical results using a NLD reduced by a factor of 5, with and without including discrete excited states. With discrete excited states, the renormalization of the NLD only affects the $\gamma$ width at higher energy, thus leading to a better reproduction of the data at higher energy but leaving the discrepancy at lower energy. A slightly better energy dependence can be obtained when applying a factor of 0.3 to the $\gamma$ width and leaving the NLD unchanged.

This raises the question of the completeness of the experimental level scheme. Discrete excited states were taken from the 2010 versions of \cite{nudat,ensdf}, up to an energy of 3.943~MeV. It is a well-known problem in Hauser-Feshbach calculations that it is advantageous, on one hand, to include low-lying excited states but, on the other hand, the included level information has to be complete to guarantee accurate predictions. It is often hard to decide at which excitation energy a cutoff should be made, especially in global calculations, in which the cutoff has to be implemented through some automated algorithm. The difficulty reproducing the energy dependence of the measured cross sections encountered here points to such a problem. For another recent case and further discussion, see \cite{i127}.

The data do not extend very much into the energy region where the proton width dominates the energy dependence. Similar to the other proton capture reactions discussed above, it seems that the proton width is well described by the optical potential of \cite{lej}. Therefore, the reactivities and their fit coefficients, shown in Tables \ref{tab:nirate} and \ref{tab:nifit}, respectively, are based on the SMARAGD calculation using this proton potential and the renormalized $\gamma$ width without inclusion of discrete excited states (shown as solid line in Fig.\ \ref{fig:ni60nld}).

\subsection{$^{58}$Ni(p,$\gamma$)$^{59}$Cu}

\begin{figure}
\includegraphics[angle=-90,width=\columnwidth]{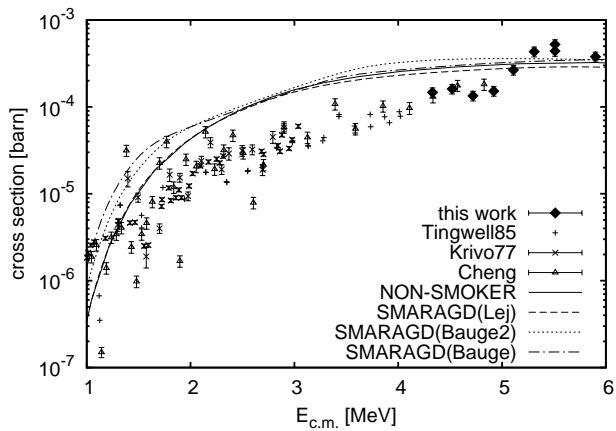}
\caption{\label{fig:ni58comp}Comparison of the reaction cross sections of $^{58}$Ni(p,$\gamma$)$^{59}$Cu to predictions. Data are from this work and \cite{kriv77} (Krivo77), \cite{tin85} (Tingwell85), \cite{che80} (Cheng).}
\end{figure}

\begin{figure}
\includegraphics[angle=-90,width=\columnwidth]{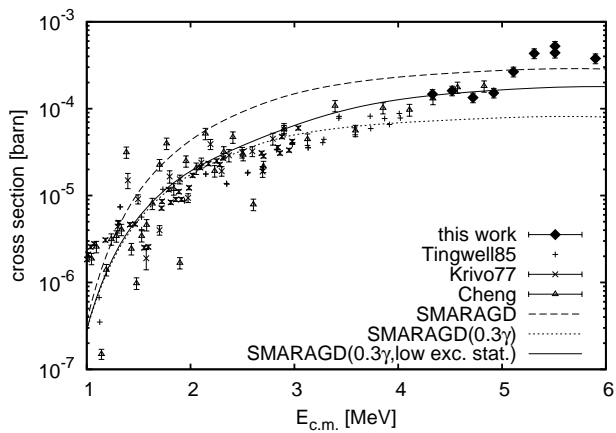}
\caption{\label{fig:ni58nld}Comparison of measured reaction cross sections of $^{58}$Ni(p,$\gamma$)$^{59}$Cu to predictions using excited states in the compound nucleus up to above the proton separation energy (SMARAGD), the same but with the $\gamma$ width reduced by a factor of 0.3 ($0.3\gamma$), and a calculation using a limited set of excited states and a reduced $\gamma$ width ($0.3\gamma$,low exc.\ stat.). See text for details. Data are from this work and \cite{kriv77} (Krivo77), \cite{tin85} (Tingwell85), \cite{che80} (Cheng).}
\end{figure}

The NLD at compound formation energy is clearly too low for this reaction to expect that all features of the cross sections can be described by an average over resonances. Moreover, we encounter a similar problem as for $^{60}$Ni(p,$\gamma$)$^{61}$Cu regarding the completeness of the included discrete level scheme of $^{59}$Cu. The proton separation energy in $^{59}$Cu is $S_\mathrm{p}=3.419$~MeV. The calculations of NON-SMOKER and SMARAGD, shown in Fig.\ \ref{fig:ni58comp}, made use of 19 levels ($E_\mathrm{max}=3.1141$~MeV \cite{adndt1}) and of 29 levels ($E_\mathrm{max}=4.307$~MeV) above the g.s., respectively. The levels included below $S_\mathrm{p}$ are very similar, despite of the fact that SMARAGD is using a more recent version of \cite{nudat,ensdf}. Both codes yield almost the same results, with a very similar energy dependence which is different from the experimentally found one.

Similar to the $^{60}$Ni case discussed above, better agreement between calculation and experiment can be achieved by including fewer excited states in the calculation. Figure \ref{fig:ni58nld} shows how a combination of a $\gamma$ width reduced by a factor of 0.3 and a limitation of the level set to excitation energies $E_\mathrm{exc}\leq 2.391$~MeV (the $J^\pi=9^-$ level) reproduces the mean data over a wide range of energies. Considering only the g.s.\ and neglecting all excited states gives the same result, indicating that the NLD is predicted well in this confined energy range. Using the same $\gamma$ width renormalization with the full level scheme results in a too low cross section at higher energies, indicating the incompleteness of the level scheme.

The data by \cite{kriv77,tin85,che80} and from this work agree with each other well although the cross section variation structures seem to be different. It has to be realized, however, that no high-resolution experiment was performed, able to resolve the (partially overlapping) individual resonances expected for this reaction. Depending on the exact beam energy and beam profile, different parts of the same resonances are sampled, leading to seemingly different patterns. Therefore, the larger scatter seen in the data by \cite{che80} may well be compatible with the other data. As mentioned in Sec.\ \ref{sec:theorygeneral}, a statistical Hauser-Feshbach calculation can only aim to reproduce the average cross sections, which seems to be successfully done by the renormalized calculation shown in Fig.\ \ref{fig:ni58nld}. For the same reason, however, it may be misleading to use the experimental data points to compute a reaction rate. They do not fully resolve the resonances and a simple interpolation will lead to incorrect results. Using the theoretical, already averaged values will yield a more realistic rate, when lacking further knowledge of the resonance properties. Therefore, the reactivities provided in Table \ref{tab:nirate} are based on the calculation using the renormalized $\gamma$ width and the limited set of excited states (shown as solid line in Fig.\ \ref{fig:ni58nld}). The REACLIB fit coefficients for these reactivities are given in Table \ref{tab:nifit}.

\section{Conclusions for $^{56}$N\lowercase{i}(\lowercase{p},$\gamma$)$^{57}$C\lowercase{u}}
\label{sec:ni56pg}

\begin{figure}
\includegraphics[angle=-90,width=\columnwidth]{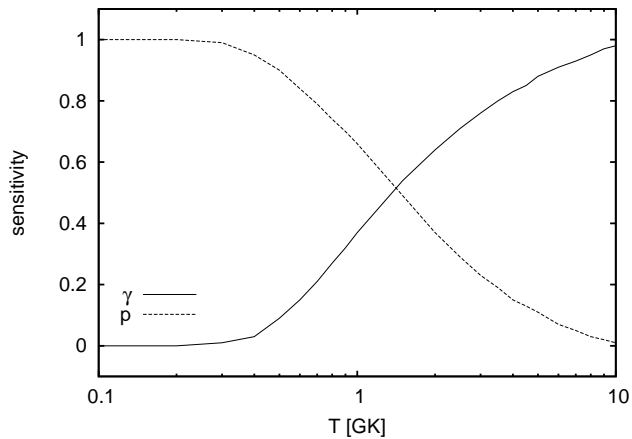}
\caption{\label{fig:sensiNi56}Sensitivity of the stellar reactivity of $^{56}$Ni(p,$\gamma$)$^{57}$Cu to variations of $\gamma$ and particle widths, shown as function of plasma temperature $T$.}
\end{figure}

One has to be careful in drawing conclusions for the stellar $^{56}$Ni(p,$\gamma$)$^{57}$Cu rate from the above trends. It is not clear to what extent the conclusions can be applied to the doubly magic nucleus with a very low proton separation energy of 0.695~MeV. Its reactivity depends on both the proton and $\gamma$ widths down to 1 GK, as shown in Fig.\ \ref{fig:sensiNi56}. Due to the low $Q$ value, however, it will equilibrate with its reverse rate already at lower temperature than most other proton captures in the $\nu p$-process path.

In the discussion above it was found that the proton widths were predicted well in all cases. A renormalization of the $\gamma$ widths by a factor of 0.3 (0.25 for $^{64}$Ni(p,$\gamma$)$^{65}$Cu) was required in all cases, except for $^{62}$Ni(p,$\gamma$)$^{63}$Cu. The data for the latter reaction was reproduced well by the SMARAGD prediction but slightly increasing deviations at higher energy may indicate that a suppression of the $\gamma$ width may be required, too. Problems with the NLD only arose from incomplete level schemes of low-lying states, not from theoretical NLDs.

\begin{table}
\caption{Predicted reactivity of $^{56}$Ni(p,$\gamma$)$^{57}$Cu as function of plasma temperature $T$.\label{tab:ni56rate}}
\begin{ruledtabular} \begin{tabular}{cl}
\multirow{2}{*}{$T$ [GK]}	& $N_\mathrm{A} \left< \sigma v \right>$ \\
						& [cm$^3$ s$^{-1}$ mole$^{-1}$] \\
\hline
 0.10 & $  9.665\times 10^{-21}$ \\
 0.15 & $  3.243\times 10^{-16}$ \\
 0.20 & $  2.154\times 10^{-13}$ \\
 0.30 & $  6.990\times 10^{-10}$ \\
 0.40 & $  9.798\times 10^{-8}$ \\
 0.50 & $  2.758\times 10^{-6}$ \\
 0.60 & $  3.058\times 10^{-5}$ \\
 0.70 & $  1.895\times 10^{-4}$ \\
 0.80 & $  7.983\times 10^{-4}$ \\
 0.90 & $  2.571\times 10^{-3}$ \\
 1.00 & $  6.811\times 10^{-3}$ \\
 1.50 & $  1.669\times 10^{-1}$ \\
 2.00 & $  1.032$ \\
 2.50 & $  3.445$ \\
 3.00 & $  8.269$ \\
 3.50 & $  1.628\times 10^{1}$ \\
 4.00 & $  2.821\times 10^{1}$ \\
 4.50 & $  4.473\times 10^{1}$ \\
 5.00 & $  6.653\times 10^{1}$ \\
 6.00 & $  1.286\times 10^{2}$ \\
 7.00 & $  2.179\times 10^{2}$ \\
 8.00 & $  3.319\times 10^{2}$ \\
 9.00 & $  4.543\times 10^{2}$ \\
10.00 & $  5.533\times 10^{2}$
\end{tabular} \end{ruledtabular}
\end{table}

\begin{table*}
\caption{Fit parameters (in REACLIB format \cite{adndt}) for the reactivity of $^{56}$Ni(p,$\gamma$)$^{57}$Cu and its reverse reaction. The reverse value has to be multiplied by the ratio of the partition functions to obtain the ($\gamma$,p) reactivity.\label{tab:ni56fit}}
\begin{ruledtabular} \begin{tabular}{cccccccc}
&$a_0$ 		& $a_1$ 			&$a_2$ 			& $a_3$ 			& $a_4$ 		& $a_5$ 			& $a_6$\\
\hline
&  $7.278475\times 10^{1}$&  0.000000& $-4.574355\times 10^{1}$& $-3.572965\times 10^{1}$&  $3.975448$& $-2.806210\times 10^{-1}$&  $2.177780$ \\
rev&  $9.507767\times 10^{1}$& $-8.064315$& $-4.574355\times 10^{1}$& $-3.572965\times 10^{1}$&  $3.975448$& $-2.806210\times 10^{-1}$&  $3.677780$
\end{tabular} \end{ruledtabular}
\end{table*}

Applying the above, an educated guess for an improved $^{56}$Ni(p,$\gamma$)$^{57}$Cu reactivity can come from a SMARAGD calculation using the proton potential by \cite{lej}, a renormalized $\gamma$ width, and no discrete excited states. It has to be emphasized once again that such a calculation can only yield an average value for the cross sections which will only be a good approximation for obtaining the reactivity if, folded with a Maxwell-Boltzmann distribution, they yield a similar value as the actual resonance contributions. This will not be the case if only few, widely separated resonances contribute to the reaction rate integral \cite{wor94}.

\begin{figure}
\includegraphics[angle=-90,width=\columnwidth]{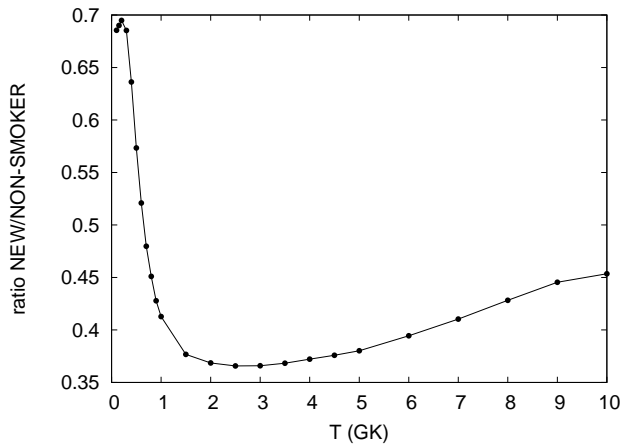}
\caption{\label{fig:ni56ratio}Ratio of the newly derived rate for $^{56}$Ni(p,$\gamma$)$^{57}$Cu and the standard NON-SMOKER rate \cite{adndt1}, shown as function of plasma temperature $T$.}
\end{figure}

Table \ref{tab:ni56rate} gives the reactivity values obtained with the above procedure for the $^{56}$Ni(p,$\gamma$)$^{57}$Cu reaction and the REACLIB fit parameters are provided in Table \ref{tab:ni56fit}. The ratio of the new rate to the previously widely used standard NON-SMOKER rate is shown in Fig.\ \ref{fig:ni56ratio}. In the relevant temperature range, the new rate is lower than the previous standard rate by a factor of about 0.37. However, the new reaction rate has a negligible impact on the final abundances predicted by the $\nu p$ process calculations due to the aforementioned equilibrium between the (p,$\gamma$) and ($\gamma$,p) reactions.

\section{Summary and conclusions}
\label{sec:sum}

In summary, the cross sections of the (p,$\gamma$) reaction for all stable nickel isotopes have been measured. The results were compared with previous data found in literature. For $^{64, 62, 58}$Ni a good agreement with previous results was found. For $^{61,60}$Ni, the new results confirmed those of \cite{tin88,tin88a} and disagree with the values of \cite{kriv77}. All the results were compared with the standard NON-SMOKER calculations and with new predictions of the SMARAGD code.
New reaction rates were proposed for all the reactions studied in the present work. In addition, a new estimate of the $^{56}$Ni(p,$\gamma$)$^{57}$Cu rate was derived based on the comparisons of predictions and data for other proton-rich Ni isotopes. The new rate is lower by a factor of about 0.37 than the previously used standard rate, but the change has no significant influence on the $\nu p$ process calculations. A generalization to all proton captures in the $\nu p$ process (or other nucleosynthesis processes), however, cannot be made as the deviations between experiment and theory are specific to the nuclides investigated here.

\section*{Acknowledgements}
This work was supported by the National Science Foundation under Grants No. PHY-1102511,  PHY-0822648 (Joint Institute for Nuclear Astrophysics), the EUROCORES EuroGENESIS research program, the ENSAR/THEXO European FP7 program, the Hungarian Academy of Science, and by the DOE Topical Collaboration "Neutrinos and Nucleosynthesis in Hot and Dense Matter" under contract DE-FG02-10ER41677.

\end{document}